\documentclass{WileyMSP-template}
\usepackage{xcolor}
\usepackage{ragged2e}
\usepackage[colorlinks=true,linkcolor=blue,citecolor=blue,urlcolor=black]{hyperref}

\usepackage[normalem]{ulem}
\usepackage{booktabs}
\usepackage{siunitx}
\usepackage{threeparttable}
\usepackage{tabularx}
\usepackage{makecell}
\usepackage{bm}
\usepackage{textcomp}
\begin{document}

\pagestyle{fancy}
\rhead{\includegraphics[width=2.5cm]{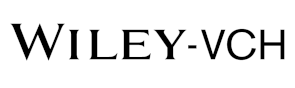}}

\title{Intense and stable second-harmonic generation in robust MePEA-based quasi-2D halide perovskite crystals}

\maketitle

\author{Grigorii~Verkhogliadov\textsuperscript{1}*,}
\author{Dmitry~Semyonov\textsuperscript{2$\dagger$},}
\author{Elizaveta~Sapozhnikova\textsuperscript{1,2$\dagger$},}
\author{Dmitry~Tatarinov\textsuperscript{1,2},}\\
\author{Marina~Tepliakova\textsuperscript{3},}
\author{Nina~Ratmanova\textsuperscript{4},}
\author{Ivan~Andreev\textsuperscript{4},}
\author{Ivan~Melnichenko\textsuperscript{5},}
\author{Natalia~Kryzhanovskaya\textsuperscript{5},}
\author{Alexey~Zhukov\textsuperscript{5},}
\author{Quynh~Le-Van\textsuperscript{6},}
\author{Daria~Khmelevskaia\textsuperscript{5}*,}
\author{Anatoly~Pushkarev\textsuperscript{1}*}

\begin{affiliations}
$^1$ Center for Photonic Science and Engineering, Skolkovo Institute of Science and Technology, Moscow, 121205, Russia.

$^2$ School of Physics and Engineering, ITMO University, St. Petersburg, 197101, Russia.

$^3$ Center for Energy Science and Technology, Skolkovo Institute of Science and Technology, Moscow, 121205, Russia.

$^4$ Department of Chemistry, M.V. Lomonosov Moscow State University, Leninskie gory 1-3, Moscow, 119991, Russia.

$^5$ International Laboratory of Quantum Optoelectronics, HSE University, St. Petersburg, 199106, Russia.

$^6$ Center of Environmental Intelligence, College of Engineering and Computer Science, VinUniversity, Gia Lam district, Hanoi, 100000, Vietnam.

$^{\dagger}$These authors contributed equally to this work.

Grigorii Verkhogliadov

Email Address: g.verkhogliadov@skoltech.ru

Daria Khmelevskaia

Email Address: dkhmelevskaya@hse.ru

Anatoly Pushkarev

Email Address: an.pushkarev@skoltech.ru

\end{affiliations}


\begin{abstract}
\justifying
Quasi-2D halide perovskites offer a unique platform for nonlinear optics applications, where structural and optical properties are tuned by the variation of organic spacer cations. Notably, quasi-2D perovskite single crystals can demonstrate intense and stable second-harmonic generation signals when a proper composition is selected. Here, the influence of phenethylammonium and 4-methyl-2-phenylethylammonium spacers on the crystal symmetry and second-harmonic signal response of quasi-2D perovskite phases with a varying number of single-octahedron [PbX$_6$]$^{4-}$ layers ($n$) is investigated. High-quality perovskite crystals with lateral dimensions of several hundred micrometers are fabricated using a cavity-confined crystallization synthesis method. Optical characterization of the crystals shows strong and temporally stable second-harmonic generation. Power-dependent and polarization-resolved measurements confirm the second-order nonlinear origin of the observed response and reveal anisotropic emission patterns. Our results demonstrate a direct correlation between spacer-induced center-symmetry breaking and nonlinear optical activity. These results establish quasi-2D halide perovskites as a promising class of symmetry-engineered materials for efficient frequency conversion and integrated nonlinear photonics.
\end{abstract}

\section{Introduction}
\justifying
Layered quasi-two-dimensional (quasi-2D) halide perovskites represent a promising class of hybrid organic-inorganic semiconductor materials for next-generation optoelectronics~\cite{zhou2022spacer,gu2025quasi} and nonlinear photonics~\cite{wang2021advances,dong2024external}. They consist of halide perovskite layers separated by insulating organic spacer layers. The band gap of the latter is larger than that of the former, which results in the localization of the excited charge carriers in the perovskite layers, i.e. reproducing multiple quantum wells~\cite{leung2022stability}. Therefore, optical and electronic properties of these materials are dependent on the quantum confinement and dielectric screening effects, which, in turn, are governed by the number of single-octahedron [MX$_6$]$^{4-}$ (M = Pb$^{2+}$, Sn$^{2+}$; X = Cl$^-$, Br$^-$, I$^-$) layers ($n$) in A$'_2$A$_{n-1}$M$_n$X$_{3n+1}$ (A$'$ - organic spacer; A = MA$^+$, FA$^+$, Cs$^+$) structure~\cite{sichert2015quantum,polavarapu2017advances}. 

\par With increasing $n$, the exciton energy decreases due to weaker quantum confinement, while the electron-hole binding energy also decreases as a result of enhanced dielectric screening~\cite{zhang2017structures}. Consequently, the optical absorption edge undergoes a redshift, the exciton diffusion length increases, and the fraction of free charge carriers contributing to electrical conductivity gradually rises. In addition to the aforementioned tunability, quasi-2D perovskites have been demonstrated to exhibit enhanced moisture stability, suppressed ion migration, and lower structural defect densities compared with their bulk counterparts~\cite{leung2022stability,laxmi2025optical}. Together, these characteristics position quasi-2D perovskites as a reliable bridge between strictly 2D and bulk semiconductors.

Beyond their linear optical properties, quasi-2D perovskites are attracting increasing attention as nonlinear optical (NLO) materials~\cite{renpolarization,fu2025ultrafast}. Among various NLO processes, second-harmonic generation (SHG) is of particular importance due to its wide range of photonic applications, including frequency conversion in laser systems~\cite{lu2024emerging}, high-resolution imaging~\cite{zhang2024polarization,schlickriede2020nonlinear}, and optical sensing~\cite{mesch2016nonlinear}. SHG is governed by the second-order nonlinear susceptibility tensor $\chi^{(2)}$. Within the electric-dipole approximation, this tensor is strictly forbidden in centrosymmetric crystals~\cite{chen2021high}. Consequently, efficient SHG can only occur in materials with a noncentrosymmetric crystal structure.

In layered halide perovskites, the incorporation of organic spacers can introduce structural asymmetry that is absent in bulk perovskites, resulting from octahedral distortions and spacer ordering. Hence, these materials are highly attractive for symmetry-engineered nonlinear optics due to proper spacer selection~\cite{chakraborty2023rational,mihalyi2024revealing}. Until now, most SHG-active quasi-2D perovskites have been leveraged linear alkyl spacers such as butylammonium (BA) halide~\cite{li2019two,stoumpos2016ruddlesden} or aromatic spacers like phenethylammonium (PEA) halide~\cite{morrow2020disentangling}. Nevertheless, BA- and PEA-based quasi-2D perovskites demonstrate poor long-term stability and low SHG conversion efficiency, which limits their potential practical application~\cite{morrow2020disentangling,li2022stereochemically}. Alternatively, chiral organic spacers provide another promising way to form noncentrosymmetric crystal structures with strong SHG~\cite{guo2023spatially,li2025regulating, yu2026second}. However, the synthesis complexity of chiral ligands and their frequently limited structural and operational stability is considered as a significant issue for scalable nonlinear photonics~\cite{ma2021recent,yang2025technical}. Consequently, rational spacer design strategies is critical for achieving simultaneously high SHG efficiency and structural robustness~\cite{yao2025stable,ma2021first}. In this context, methyl-substituted phenethylammonium spacers provide an attractive option for symmetry engineering without introducing molecular chirality. Specifically, the additional methyl group in 4-methyl-2-phenylethylammonium (MePEA) can modify molecular packing, induce lattice distortions, and afford noncentrosymmetry in quasi-2D perovskite structures. Although MePEA ligand has already demonstrated an excellent performance as a spacer cation in perovskite solar cells~\cite{duong2023bulk,chang2024enhanced,degani2025compositional} and blue light-emitting diodes~\cite{alahbakhshi2023highly}, its potential to enable nonlinear optical functionality in quasi-2D perovskites remains unexplored.

Here, we systematically study bromide quasi-2D perovskite single crystals with $n=1-3$ PEA and MePEA spacer cations. Bromide compositions were selected due to their enhanced resistance to laser-induced degradation and improved stability compared to iodide-based crystals~\cite{aziz2019understanding,mcgovern2020understanding,ren2025elucidation}. We obtain high-quality single crystals via the cavity-confined crystallization method and directly study centrosymmetric and noncentrosymmetric perovskite phases. Through power-dependent and polarization-resolved SHG measurements, we establish a clear structure-property relationship between spacer-induced structural diversity and nonlinear optical response and demonstrate spacer-driven crystal symmetry engineering as an effective route toward robust second-order nonlinear functionality in quasi-2D halide perovskites.

\section{Results and Discussion}
\subsection{Growth and characterization of quasi-2D perovskite microcrystals}
\begin{figure}[t!]
  \includegraphics[width=\linewidth]{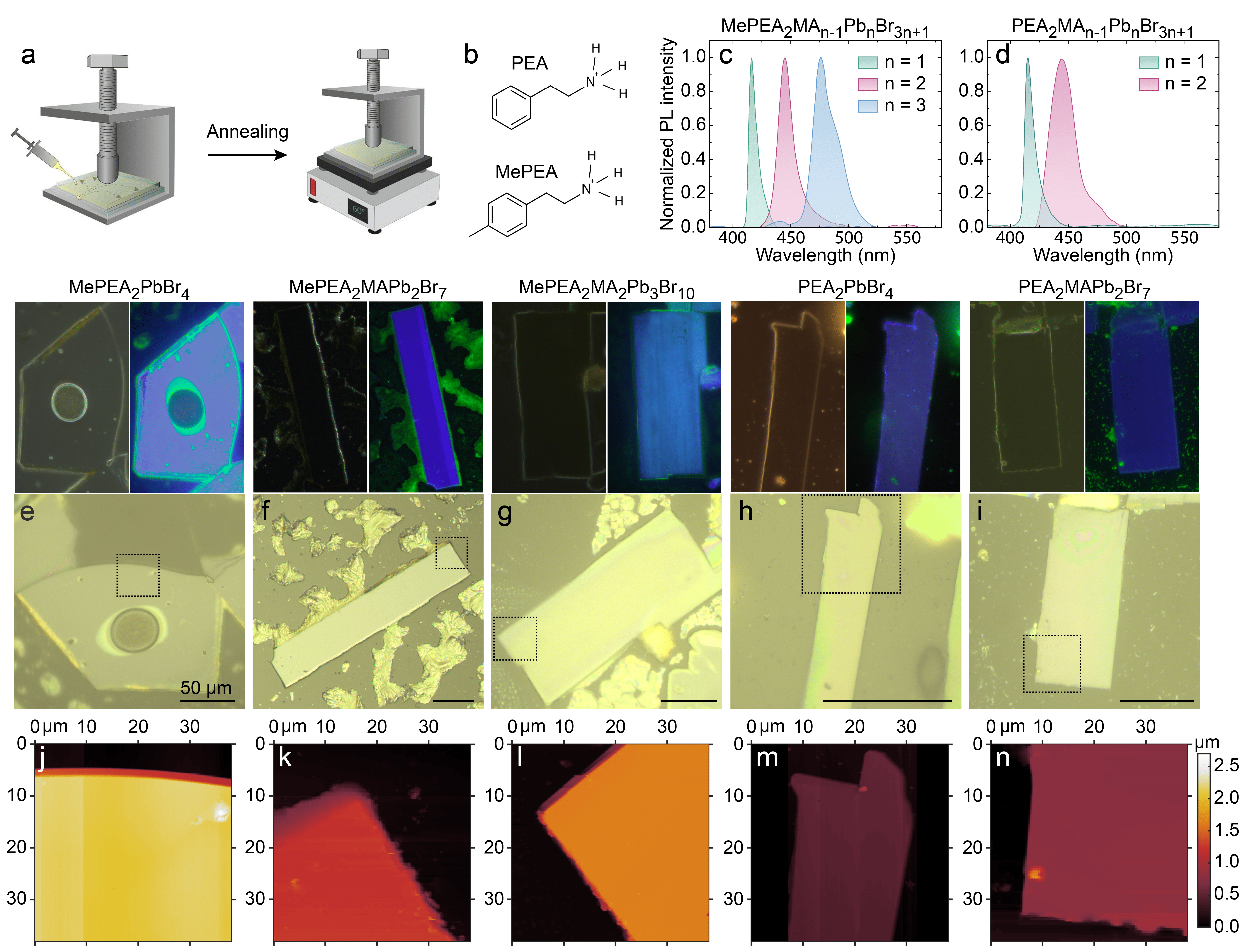}
  \caption{\textbf{Fabrication and characterization of quasi-2D perovskite crystals.}
		(\textbf{a}) Schematic illustration of the cavity-confined crystallization process.
(\textbf{b}) Chemical structures of the PEA and MePEA organic spacer cations.
(\textbf{c},\textbf{d}) PL spectra of MePEA$_2$MA$_{n-1}$Pb$_n$Br$_{3n+1}$ ($n=1$, 2, and 3) and PEA$_2$MA$_{n-1}$Pb$_n$Br$_{3n+1}$ ($n=1$ and 2) crystals, respectively.
(\textbf{e}\textbf{--}\textbf{i}) Bright-field optical microscopy images (lower panel), dark-field (upper-left panel) and PL (upper-right panel) images of MePEA$_2$PbBr$_4$ (\textbf{e}), MePEA$_2$MAPb$_2$Br$_7$ (\textbf{f}),  MePEA$_2$MA$_2$Pb$_3$Br$_{10}$ (\textbf{g}),  PEA$_2$PbBr$_4$ (\textbf{h}), and  PEA$_2$MAPb$_2$Br$_7$ (\textbf{i}) crystals. Scale bars, 50~$\mu$m. (\textbf{j}\textbf{--}\textbf{n}) AFM images of the regions highlighted in (\textbf{e}\textbf{--}\textbf{i}).
}
  \label{fig:1}
\end{figure}

To fabricate quasi-2D perovskite crystals, we employed a cavity-confined crystallization method. As schematically illustrated in Figure~\ref{fig:1}a, the precursor solution was confined within a cavity formed by two glass substrates and subsequently annealed at 60$^{\circ}$C to induce crystal growth, as described in the Materials and Methods section. This approach enables the growth of high-quality quasi-2D perovskite crystals with lateral dimensions of several hundred micrometers and thicknesses of up to several micrometers. In this work, we focused on quasi-2D perovskites incorporating phenethylammonium (PEA) and 4-methylphenethylammonium (MePEA) as the organic spacer cations. The molecular structures of these spacer cations differ only by the presence of an additional methyl substituent at the para (4-) position of the phenyl ring in MePEA, as illustrated in Figure~\ref{fig:1}b.

We synthesized five quasi-2D perovskite compositions using this method: MePEA$_2$PbBr$_4$, MePEA$_2$\allowbreak MAPb$_2$Br$_7$, MePEA$_2$MA$_2$Pb$_3$Br$_{10}$, PEA$_2$PbBr$_4$, and PEA$_2$MAPb$_2$Br$_7$, corresponding to MePEA-based phases with $n=1, 2, 3$ and PEA-based phases with $n=1,2$, respectively. The photoluminescence (PL) spectra of these crystals (Figures~\ref{fig:1}c and \ref{fig:1}d) exhibit well-defined emission peaks that systematically blueshift with decreasing $n$, reflecting the stronger quantum confinement in lower-$n$ quasi-2D perovskites. The absence of additional emission bands confirms the formation of predominantly phase-pure quasi-2D structures, whereas the weak shoulder observed for the MePEA-based $n=3$ crystal is attributed to a minor amount of the $n=4$ phase, whose formation at the crystal edges and surface is promoted by humid environmental conditions \cite{leung2020mixed,zou2018efficient,stoumpos2017high}. Bright-field (lower panel), dark-field (upper-left panel), and PL images (upper-right panel) in Figures~\ref{fig:1}e--i demonstrate well-defined crystal morphology and spatially uniform photoluminescence, indicative of high crystal quality. AFM measurements further reveal smooth crystal surfaces with a root-mean-square (RMS) roughness of around 4~nm and crystal thicknesses ranging from 0.3 to 2.1~$\mu$m (Figures~\ref{fig:1}j--n).
Representative crystals investigated in this work have thicknesses of 2.1 $\mu$m for MePEA$_2$PbBr$_4$, 1.1 $\mu$m for MePEA$_2$MAPb$_2$Br$_7$, 1.5 $\mu$m for MePEA$_2$MA$_2$Pb$_3$Br$_{10}$, 0.4 $\mu$m for PEA$_2$PbBr$_4$, and 0.8 $\mu$m for PEA$_2$\allowbreak MAPb$_2$Br$_7$. 

Although the investigated crystals exhibit high structural quality and sufficient thickness for SHG measurements, crystal symmetry remains the key factor governing the nonlinear optical response. PEA-based quasi-2D perovskites, including PEA$_2$PbBr$_4$ and PEA$_2$MAPb$_2$Br$_7$, are reported to crystallize in the centrosymmetric $P\bar{1}$ space group~\cite{zhang2019pressure,leung2019quasi} and are therefore not expected to exhibit measurable SHG.

\begin{figure}[t!]
  \includegraphics[width=\linewidth]{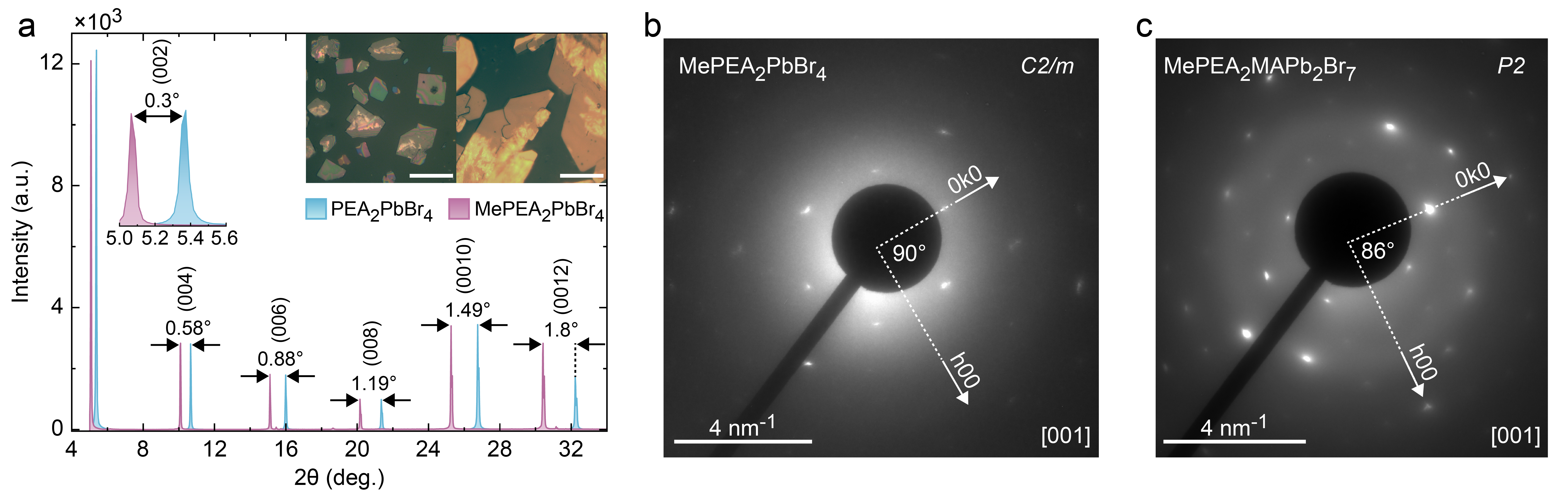}
  \caption{\textbf{Structural characterization of quasi-2D perovskite crystals.} (\textbf{a}) XRD pattern of PEA$_2$PbBr$_4$ and MePEA$_2$PbBr$_4$ crystals. Insets: enlarged view of the (002) reflections and optical images of the PEA$_2$PbBr$_4$ and MePEA$_2$PbBr$_4$ crystals. Scale bars, 50~$\mu$m. (\textbf{b,c}) SAED patterns of the MePEA$_2$PbBr$_4$ and MePEA$_2$MAPb$_2$Br$_7$ crystals, respectively.}
  \label{fig:2}
\end{figure}

We therefore investigated the effect of methyl substitution on the crystal structure using X-ray diffraction (XRD). The XRD patterns of PEA$_2$PbBr$_4$ and MePEA$_2$PbBr$_4$ are similar, with the diffraction peaks of MePEA$_2$PbBr$_4$ systematically shifted toward lower diffraction angles, indicating an increase in the interlayer spacing between adjacent inorganic layers (Figure~\ref{fig:2}a). Specifically, the (002) reflection shifts from 5.37$^{\circ}$ for PEA$_2$PbBr$_4$ to 5.07$^{\circ}$ for MePEA$_2$PbBr$_4$, corresponding to an increase in the interlayer spacing from 16.43 to 17.43~\r{A}. Despite the incorporation of two additional methyl groups, each with an approximate size of 2~\r{A}~\cite{hihath2012effects}, this increase amounts to only about 1~\r{A}. This observation suggests a reorientation of the organic spacer molecules, resulting in modified molecular packing and associated structural distortions that may alter the crystal symmetry.

To examine the crystal symmetry of the MePEA-based quasi-2D perovskites, we performed selected area electron diffraction (SAED) measurements. The SAED patterns of MePEA$_2$PbBr$_4$ and MePEA$_2$MAPb$_2$Br$_7$ single crystals are shown in Figures~\ref{fig:2}b and \ref{fig:2}c, respectively.
The SAED pattern of MePEA$_2$PbBr$_4$ reveals a centered rectangular 2D Bravais lattice with primitive vectors $a\neq b$ ($a=2.57$~\r{A}, $b=4.27$~\r{A}) and $\gamma=90^{\circ}$ ($a\parallel h$ and $b\parallel k$)~\cite{lechthaler2020objective}. Such a lattice is compatible with several possible space groups. Considering that related layered perovskites are commonly reported to crystallize in the centrosymmetric monoclinic space group $C2/m$~\cite{cortecchia2017cathodoluminescence}, MePEA$_2$PbBr$_4$ is most likely centrosymmetric, consistent with the behavior of most $n=1$ layered perovskites.
In contrast, the SAED pattern of MePEA$_2$MAPb$_2$Br$_7$ reveals an oblique 2D Bravais lattice with $a\neq b$ ($a=7.97$~\r{A}, $b=8.10$~\r{A}) and $\gamma=86^{\circ}$. This lattice geometry is compatible with low-symmetry space groups such as $P1$ and $P2$. The observed diffraction symmetry indicates the presence of a twofold rotational axis, supporting the assignment to the non-centrosymmetric space group $P2$ and indicating the absence of inversion symmetry.

\subsection{Second-harmonic generation in quasi-2D perovskites}

The structural characterization suggests that methyl substitution of the PEA spacer can promote structural distortions and inversion-symmetry breaking in quasi-2D perovskites. To investigate the impact of spacer-cation engineering on the second-order nonlinear optical response, we performed a systematic SHG study of all five quasi-2D perovskite compositions investigated in this work. The SHG measurements were performed in a reflection geometry using femtosecond laser excitation at a wavelength of 1045~nm (see Section S1 of the Supporting Information for experimental details).

\begin{figure}[t!]
  \includegraphics[width=\linewidth]{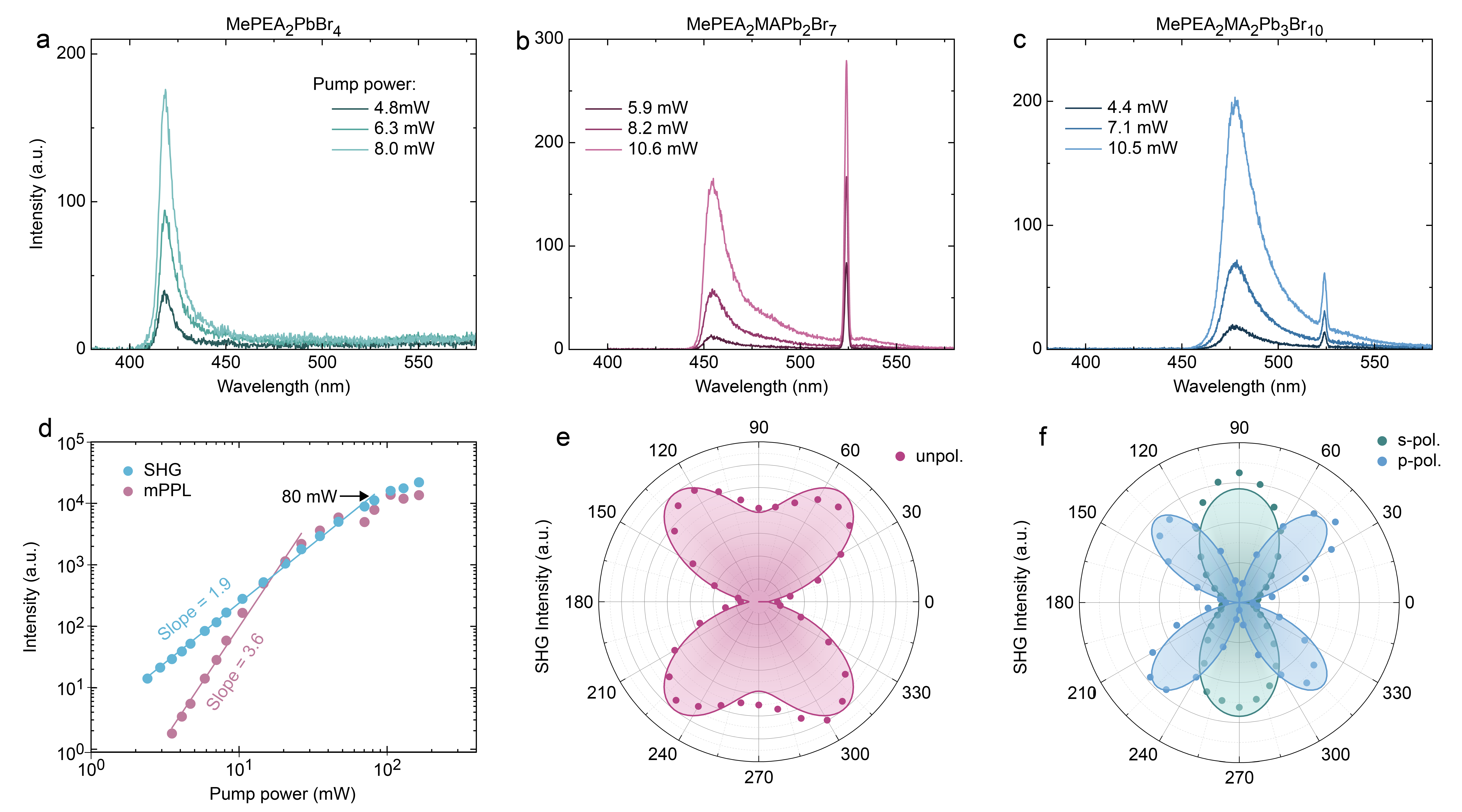}
  \caption{\textbf{SHG in MePEA-based quasi-2D perovskite crystals under excitation at $\lambda$$_{exc}$ = 1045~nm.}  (\textbf{a--c}) Nonlinear emission spectra of MePEA$_2$PbBr$_4$, MePEA$_2$MAPb$_2$Br$_7$, and MePEA$_2$MA$_2$Pb$_3$Br$_{10}$, respectively. (\textbf{d}) Pump-power dependence of the SHG and mPPL intensities of MePEA$_2$MAPb$_2$Br$_7$. (\textbf{e,f}) Polar plots of the total and polarization-resolved SHG intensities as functions of the incident polarization angle. Dots represent the experimental data, and solid lines represent the theoretical fits.}
  \label{fig:3}
\end{figure}

In agreement with the structural characterization, the nonlinear optical responses of the MePEA-based quasi-2D perovskites differ markedly. As depicted in Figure~\ref{fig:3}a, MePEA$_2$PbBr$_4$ exhibits multi-photon photoluminescence (mPPL), whereas no detectable SHG is observed prior to laser-induced damage, consistent with its likely centrosymmetric structure. Likewise, no measurable SHG is observed for PEA$_2$PbBr$_4$ or PEA$_2$MAPb$_2$Br$_7$ (Figure~S3, Supporting Information), in accordance with previous reports assigning both compositions to the centrosymmetric $P\bar{1}$ space group. 

In contrast, under similar excitation conditions, a pronounced SHG signal emerges from MePEA$_2$MAPb$_2$Br$_7$ alongside the mPPL emission, as shown in Figure~\ref{fig:3}b. The corresponding effective second-order nonlinear susceptibility, estimated using crystalline urea as a reference, is approximately  $\chi^{(2)}$ = 0.88~pm/V (see Section S2 of the Supporting Information for details). This value is comparable to that of KDP~\cite{fleming2018handbook} and is nearly twice as large as the value reported for BA$_2$MAPb$_2$Br$_7$~\cite{li2019two}.

SHG remains observable in MePEA$_2$MA$_2$Pb$_3$Br$_{10}$ (Figure~\ref{fig:3}c), although with substantially lower intensity than in MePEA$_2$MAPb$_2$Br$_7$. The weaker SHG response may be associated with changes in local structural distortions and dipole ordering as the inorganic layer thickness increases, potentially modifying the degree of inversion-symmetry breaking. A similar decrease in SHG intensity with increasing $n$ was previously reported for (BA)$_2$(MA)$_{n-1}$Pb$_n$I$_{3n+1}$ perovskites, where the observed SHG response was attributed to local symmetry-breaking effects~\cite{stoumpos2016ruddlesden}. Interestingly, a nonmonotonic dependence on $n$ was also observed for the third-order nonlinear susceptibility of these perovskites, highlighting the influence of quantum confinement on their nonlinear optical properties~\cite{saouma2017selective}. These findings underscore the complex relationship between inorganic layer thickness, crystal symmetry, and nonlinear optical response, which warrants further investigation.

We next focus on MePEA$_2$MAPb$_2$Br$_7$, which exhibits the strongest SHG response among the investigated compositions. To verify that the observed signal originates from a second-order nonlinear optical process, we measured the pump-power dependence of the SHG intensity, as shown in Figure~\ref{fig:3}d. The SHG signal remains clearly observable over a broad pump-power range and scales nearly quadratically with the incident pump power below 80~mW, yielding a fitted slope of approximately 1.9. At higher pump powers, the SHG intensity deviates from the quadratic dependence and gradually approaches saturation. Meanwhile, the mPPL pump-power dependence reveals a slope of 3.6, consistent with three-photon absorption followed predominantly by radiative excitonic recombination, with an additional contribution from free-carrier recombination~\cite{khmelevskaia2021excitonic}.

To further examine the crystal symmetry, we performed polarization-resolved SHG measurements on a MePEA$_2$MAPb$_2$Br$_7$ crystal. The excitation beam was incident normal to the crystal surface, while the in-plane crystal orientation was controlled by rotating the sample (see Section S1 and Figure~S7 of the Supporting Information). Figure~\ref{fig:3}e presents the total SHG intensity (without an analyzer in the detection path) as a function of the excitation polarization, characterized by a twofold angular dependence. The SHG signal was subsequently measured with the analyzer oriented parallel (p-polarized) or perpendicular (s-polarized) to the laboratory X-axis. As shown in Figure~\ref{fig:3}f, the p- and s-polarized SHG exhibit four- and two-lobe patterns, respectively, revealing a pronounced in-plane anisotropy of the nonlinear optical response. The polarization patterns are well reproduced by the theoretical model based on twofold rotational symmetry (see Section S3 of the Supporting Information for details), supporting the $P2$ space-group assignment inferred from the SAED measurements.

\subsection{Spectral characteristics and stability of SHG}

\begin{figure}[t!]
  \includegraphics[width=\linewidth]{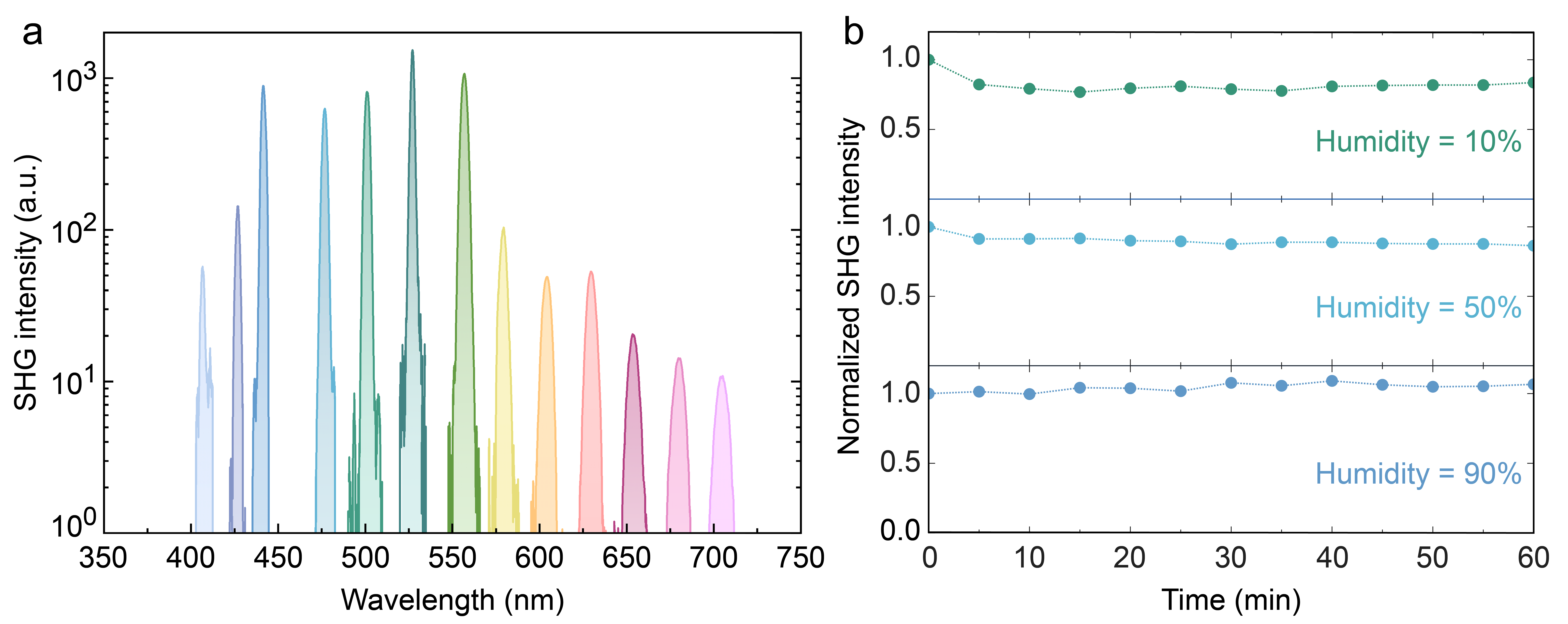}
  \caption{\textbf{Wavelength dependence and stability of SHG in MePEA$_2$MAPb$_2$Br$_7$ crystal.}
		(\textbf{a}) SHG spectra recorded under different excitation wavelengths. (\textbf{b}) Temporal evolution of the SHG intensity under continuous excitation at $\lambda_{\mathrm{exc}}=1045$~nm and a pump intensity of 9.5~GW/cm$^2$, measured at relative humidity levels of 10\%, 50\%, and 90\%.}
  \label{fig:4}
\end{figure}

We next investigate the SHG response of MePEA$_2$MAPb$_2$Br$_7$ over a broad range of excitation wavelengths (Figure~\ref{fig:4}a). As the excitation wavelength is varied from 810 to 1410~nm, the SHG peak shifts accordingly, appearing at half of the excitation wavelength. In general, the SHG intensity is expected to increase as the generated second-harmonic wavelength approaches an excitonic resonance, where the second-order nonlinear susceptibility is enhanced while absorption of the fundamental excitation remains relatively low. Accordingly, the SHG intensity increases as the second-harmonic wavelength approaches the excitonic absorption peak of the $n=2$ MePEA$_2$MAPb$_2$Br$_7$ crystal at approximately 440~nm. However, the maximum SHG intensity is observed at a wavelength of 525~nm, corresponding to a fundamental excitation wavelength of 1050~nm. The optimum SHG response therefore occurs at a longer wavelength than the excitonic absorption peak, as previously reported for quasi-2D perovskites~\cite{yuan2018chiral}. This behavior likely reflects a balance between resonance enhancement of the second-order nonlinear susceptibility and competing effects, including pump absorption, self-absorption of the generated SHG, phase mismatch, and laser-induced heating~\cite{wang2021advances,yu2022chiral}.

The temporal stability of the SHG response was initially evaluated under ambient conditions (50\% relative humidity, RH) with continuous femtosecond excitation at 1045~nm and an incident optical intensity of 9.5~GW/cm$^2$ (Figure~\ref{fig:4}b). After 60~min of irradiation, the SHG intensity retained approximately 86\% of its initial value, demonstrating sustained nonlinear optical response under prolonged excitation. By comparison, previously reported chiral (R- and S-$\beta$-methylphenethylammonium)BAPbBr$_4$ perovskites exhibited similar SHG retention of approximately 85\% after only 20~min of irradiation at a slightly lower pump intensity of 7.86~GW/cm$^2$~\cite{guo2023spatially}. Furthermore, a MePEA$_2$MAPb$_2$Br$_7$ crystal stored under ambient conditions for approximately one year continued to exhibit a pronounced SHG response that remained stable under intense femtosecond excitation (see Figure~S8 of the Supporting Information).
The effect of environmental humidity was also examined at 10\% and 90\% relative humidity using a custom-built humidity-controlled chamber equipped with a humidity sensor (see Section S1, Supporting Information). At 10\% RH, the SHG intensity retained approximately 84\% of its initial value after 60~min, whereas at 90\% RH, the response remained nearly unchanged throughout the measurement (Figure~\ref{fig:4}b). This remarkable stability under high humidity may be associated with the increased hydrophobicity of the methyl-substituted organic spacer. Methyl substitution is known to increase the lipophilicity of aromatic molecules~\cite{sun2018methyl}, potentially reducing moisture penetration into the perovskite structure. 

To further assess the resistance of MePEA$_2$MAPb$_2$Br$_7$ to intense optical excitation, we determined its laser damage threshold (LDT). The LDT was estimated from the pump intensity at which the SHG response deviates from the quadratic power dependence~\cite{yang2025chiral}. As shown in Figure~\ref{fig:3}d, this deviation occurs at a pump power of approximately 80~mW, corresponding to an LDT of 107.84~GW/cm$^2$ (16.17~mJ/cm$^2$). This value exceeds those reported for most quasi-2D and other low-dimensional halide perovskites with organic spacers under near-infrared excitation (800--1100~nm), as summarized in Table~\ref{tbl:parameters}. Under prolonged high-intensity laser irradiation, these materials can undergo photoinduced degradation associated with photooxidation~\cite{udalova2021nonmonotonic} and migration or loss of organic spacer cations~\cite{fang2018unravelling}, which can progressively disrupt the layered structure and reduce the nonlinear optical response~\cite{leung2022stability}. Resistance to such degradation is strongly influenced by the molecular structure and packing arrangement of the organic spacer. In particular, structural complexity and steric hindrance of spacer ligands can introduce additional lattice disorder and disrupt ligand packing~\cite{pols2024temperature,duan2026structure,zhang2022exploring,li2026boosting}, potentially compromising structural robustness under intense optical excitation. The high LDT of MePEA$_2$MAPb$_2$Br$_7$ may therefore be related to the improved spacer packing induced by methyl substitution, which simultaneously promotes inversion-symmetry breaking and improves resistance against laser-induced degradation, further supporting their potential for nonlinear photonic applications.

\begin{table}[htbp]
\centering
\caption{Comparison of the second-order NLO performance
of different low-dimensional perovskite materials.}
\label{tbl:parameters}

\renewcommand{\arraystretch}{1.12}

\begin{tabular*}{\linewidth}{
    @{\extracolsep{\fill}}
    l
    c c c c
    c
    c
    c
    c
    @{}
}
\toprule

& \multicolumn{4}{c}{\textbf{Laser conditions\textsuperscript{a}}}
& & & & \\[-2pt]

\cmidrule(lr){2-5}

\textbf{Material} &
\makecell[c]{$\bm{\lambda}_{\mathrm{\mathbf{ex}}}$\\\textbf{[nm]}} &
\makecell[c]{$\bm{\tau}$\\\textbf{[fs]}} &
\makecell[c]{$\bm{f}$\\\textbf{[MHz]}} &
\makecell[c]{$\bm{d}$\\\textbf{[$\bm{\mu}$m]}} &
\makecell[c]{\textbf{Power}\\\textbf{[mW]}} &
\makecell[c]{\textbf{LDT}\\\textbf{[GW cm$^{-2}$]}} &
\makecell[c]{$\bm{\chi}^{\bm{(2)}}$\\\textbf{[pm V$^{-1}$]}} &
\textbf{Ref.} \\

\midrule

(R/S-BrsBA)$_2$PbBr$_4$ &
840 & 130 & 76 & 3 &
6 & 8.59\textsuperscript{b} & - & \cite{yu2026second} \\

(S-2-C$_5$H$_{14}$N$_2$)$_2$PbI$_6$ &
960 & 100 & 80 & 20 &
745 & 29.66 & - & \cite{jia2022methylpiperazine} \\

(S-2-MPD)PbBr$_3$ &
980 & 100 & 80 & 20 &
714 & 28.42 & 29.9 & \cite{zheng20221d} \\

(R-MBA)CuBr$_2$ &
940 & 100 & 82 & - &
- & 17 & 24.7 & \cite{ge2022chiral} \\

(S-FMBA)$_2$SnCl$_6$ &
860 & 100 & 80 & 20 &
1000 & 39.81 & 18.84 & \cite{han2023linear} \\

(R-MPEA)$_6$InCl$_9$ &
800 & 100 & 80 & 20 &
961 & 38.26 & - & \cite{qi20220d} \\

DMA$_4$[InCl$_6$]Br &
1040 & 100 & 80 & 20 &
800 & 31.85 & - & \cite{guan2023free} \\

R-2-mpipSnBr$_6$ &
1000 & 100 & 80 & 20 &
1185 & 47.17 & - & \cite{yu2023oxidation} \\

(R-3-APD)SbCl$_5$ &
1020 & 100 & 80 & 20 &
961 & 39.4 & - & \cite{qi2022organic} \\

MPEA$_2$SnBr$_6$ &
840 & 100 & 80 & 20 &
420 & 16.72 & 0.33 & \cite{zhao2021tin} \\

(C$_4$H$_{10}$NO)PbBr$_3$ &
800 & 140 & 80 & 2 &
16.47\textsuperscript{b} & 46.82 & 0.32 & \cite{shen2022c4h10no} \\

(S-NEA)$_3$Ge$_2$I$_7$·H$_2$O &
880 & 100 & 80 & 6 &
86.99\textsuperscript{b} & 38.46 & 0.86 & \cite{wang2023chiral} \\

(L-Pro)PbI$_3$·H$_2$O &
920 & 100 & 80 & 20 &
700 & 27.87 & 9.5 & \cite{xin2023resonant} \\

(S-APD)PbBr$_4$ &
800 & 100 & 80 & 20 &
1491 & 59.36 & 0.4 & \cite{yang2025chiral} \\

(R-2-EP)PbBr$_4$ &
860 & 100 & 80 & 20 &
1362 & 54.22 & 0.33 & \cite{yang2025high} \\

(R-BrMBA)$_2$GeI$_4$ &
1010 & 100 & 80 & 6 &
157.7 & 69.75\textsuperscript{b} & 0.4 & \cite{fu2025competitive} \\

\textbf{MePEA$_2$MAPb$_2$Br$_7$} &
\textbf{1045} &
\textbf{150} &
\textbf{70} &
\textbf{3} &
\textbf{80} &
\textbf{107.84} &
\textbf{0.88} &
\makecell[c]{\textbf{This}\\\textbf{work}} \\

\bottomrule
\end{tabular*}

\vspace{4pt}
\begin{minipage}{\linewidth}
\footnotesize
\textsuperscript{a} $\lambda_{\mathrm{ex}}$ denotes the excitation wavelength,
$\tau$ the pulse duration, $f$ the repetition rate, and $d$ the laser spot diameter.\\
\textsuperscript{b} Calculated from the reported parameters using
$LDT=4P/(f\pi d^2\tau)$, where $P$ is the incident power at the damage
threshold.
\end{minipage}

\end{table}

\section{Conclusion}

We demonstrated that molecular engineering of an organic spacer cation provides an effective strategy for achieving robust second-harmonic generation (SHG) in quasi-2D bromide perovskites. While PEA-based and the $n=1$ MePEA-based microcrystals exhibited no detectable SHG, methyl substitution enabled a pronounced nonlinear optical response in the $n=2$ MePEA$_2$MAPb$_2$Br$_7$ crystals, with a weaker response observed for the $n=3$ counterparts. Polarization-resolved SHG measurements supported the assignment of the $n=2$ phase to the noncentrosymmetric $P2$ space group, highlighting the role of spacer-induced inversion-symmetry breaking. Notably, MePEA$_2$MAPb$_2$Br$_7$ exhibited an effective second-order nonlinear susceptibility of $\chi^{(2)} = 0.88$ pm/V, broadband SHG under excitation from 810 to 1410 nm, and an estimated laser damage threshold of 107.84 GW/cm$^2$. Its nonlinear response remains stable under prolonged femtosecond excitation at 90$\%$ relative humidity, while pronounced SHG persists after 12 months of ambient storage. Our findings reveals how a subtle molecular modification - the introduction of a single methyl group into the organic spacer cation - significantly influences crystal symmetry, nonlinear optical response, and environmental stability of quasi-2D halide perovskites. This structure-property relationship highlights organic-spacer engineering as a promising strategy for developing stable and efficient quasi-2D perovskites for optical frequency conversion and integrated nonlinear photonics.

\section{Experimental Section}
\textbf{Quasi-2D perovskite solution preparation}
\\
Quasi-2D perovskite precursor solutions of MePEA$_2$MAPb$_2$Br$_7$ and PEA$_2$MAPb$_2$Br$_7$ were prepared by dissolving 0.3~mmol PbBr$_2$ ($\geq$99.9\%, Lanhit), 0.15~mmol MABr ($\geq$99.99\%, Greatcell Solar Materials), and 0.3~mmol MePEABr ($\geq$98\%, Aladdin Scientific) or PEABr ($\geq$98\%, Greatcell Solar Materials) in anhydrous dimethyl sulfoxide (DMSO).

The MePEA$_2$PbBr$_4$ and PEA$_2$PbBr$_4$ solutions were prepared by dissolving 0.3~mmol PbBr$_2$ ($\geq$99.9\%, Lanhit) and 0.6~mmol MePEABr ($\geq$98\%, Aladdin Scientific) or PEABr ($\geq$98\%, Greatcell Solar Materials) in anhydrous DMSO. For MePEA$_2$MA$_2$Pb$_3$Br$_{10}$, 0.3~mmol PbBr$_2$ ($\geq$99.9\%, Lanhit), 0.2~mmol MABr ($\geq$99.99\%, Greatcell Solar Materials), and 0.2~mmol MePEABr ($\geq$98\%, Aladdin Scientific) were dissolved in anhydrous DMSO.
\\
\\
\textbf{Cavity-confined crystallization of perovskite microcrystals}\\
Cavity-confined crystallization (a modified space-confined crystallization method~\cite{he2019oriented}) was used to obtain crystals up to about 100~$\mu$m in size on the glass substrate. In this method, two glass substrates were attached together using a U-shaped metal bracket with a screw, the latter fitted with a cup nut, and 3~$\mu$L of the stoichiometric quasi-2D perovskite DMSO solution was then deposited between the two glass substrates, as illustrated in Figure~\ref{fig:1}a. The uniform spreading of the solution is attributed to the capillary effect. The metal bracket with substrates inside was annealed in a laboratory oven at 60$^{\circ}$C for 48 hours to induce crystal growth.
\\
\\
\textbf{Structural and optical measurements}\\
To characterize the crystal structure of the quasi-two-dimensional halide perovskite with the composition MePEA$_n$MA$_{n-1}$Pb$_n$Br$_{3n+1}$ (phases $n=1$ and $n=2$), selected-area electron diffraction patterns were recorded along a specific crystallographic zone using a ZEISS LIBRA 200FE transmission electron microscope.

For diffraction analysis, individual perovskite flakes were transferred onto 300-mesh copper grids with a carbon film. These flakes were prepared via mechanical exfoliation using Kapton tape, then transferred onto a polydimethylsiloxane (PDMS) lens. The lens was subsequently dissolved in hexane, releasing the flakes into suspension, which was then drop-cast onto the copper grids.
Suitable samples for further investigation were identified on the grids using an Axio Imager A2m optical microscope (Carl Zeiss SMT) equipped with a 350~nm UV excitation source, coupled with an Ocean Optics QE PRO spectrometer.

X-ray diffraction patterns were collected using a Bruker D8 Advance diffractometer (Bruker AXS, Germany) equipped with Cu K$\alpha$ radiation ($\lambda$ = 1.5406~\r{A}), operating at 40~kV and 40~mA.

Dark-field, bright-field, and photoluminescence images of perovskite crystals were obtained on a Karl Zeiss AxioImager A2m microscope coupled with halogen and mercury lamps utilizing 50x and 20x magnification objectives. PL spectra were measured using a QE Pro Ocean Optics spectrometer connected to the Karl Zeiss AxioImager A2m optical microscope with optical fiber.  

A detailed description of the SHG measurements, including a schematic illustration of the optical setup, is provided in Section S1 of the Supporting Information.

\section*{Supporting Information}\par 
Supporting Information is available from the Wiley Online Library or from the author.

\section*{Acknowledgements} \par
This work was supported by the Russian Science Foundation (project No. 25-22-00278,~\url{ https://rscf.ru/project/25-22-00278/}). The SAED investigations were conducted at the "Interdisciplinary Resource Center for Nanotechnology" of the Research Park of St. Petersburg State University. The authors thank D.V. Danilov for performing the SAED measurements.

\section*{Conflict of Interest} \par
The authors declare no conflict of interest.

\section*{Data Availability Statement} \par
The data that support the ﬁndings of this study are available from the corresponding author upon reasonable request.

\medskip

\bibliographystyle{MSP}
\bibliography{3.references}

\newpage

\begin{figure}
\centering
\textbf{Table of Contents}\\

\medskip
\includegraphics[width=0.3\linewidth]{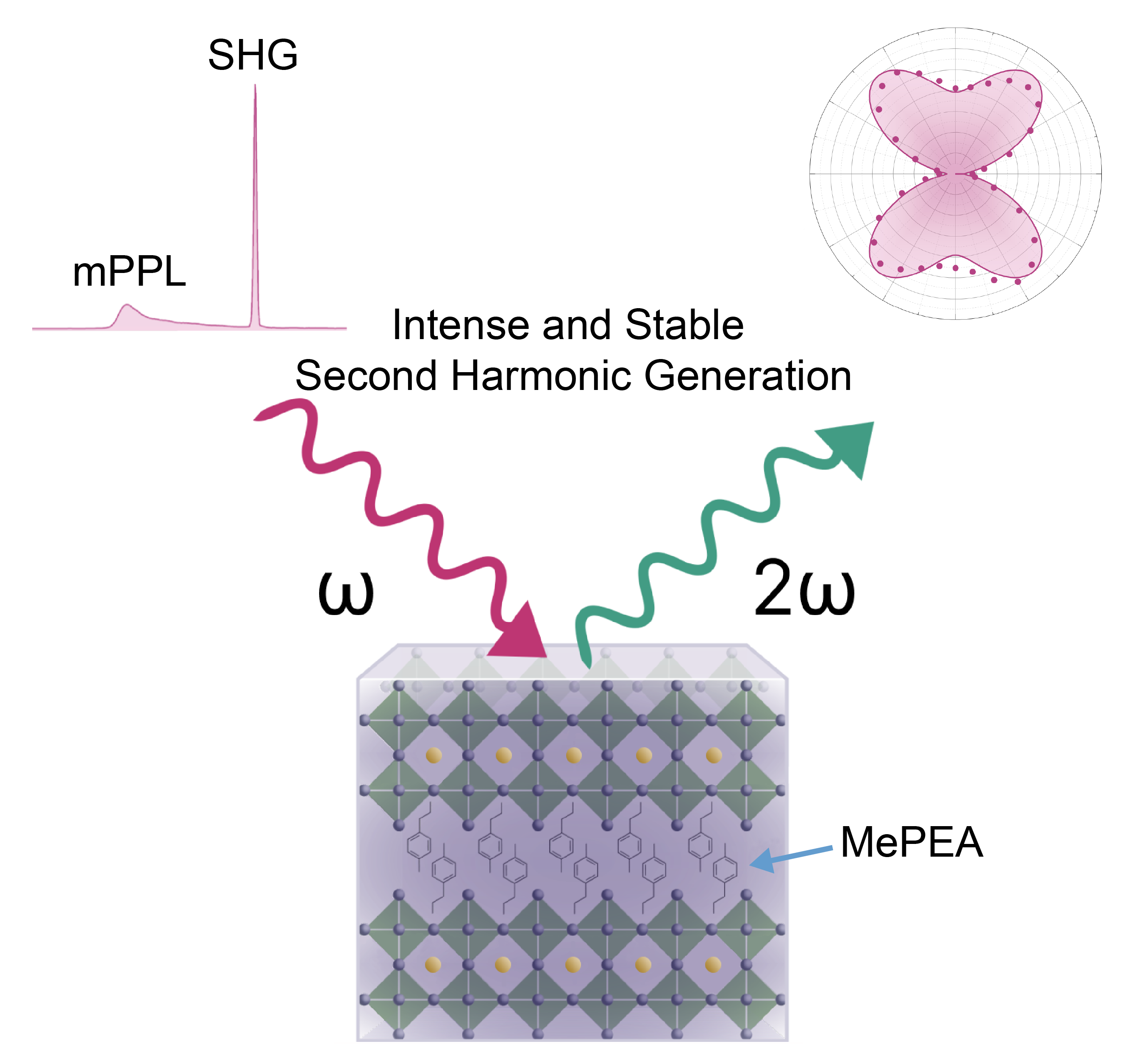}
\medskip
\caption*{The crystal symmetry and nonlinear optical response of quasi-2D bromide perovskite are controlled through molecular spacer design. Methyl-substituted phenethylammonium induces inversion-symmetry breaking and enables strong and stable second-harmonic generation. These findings demonstrate that robust MePEA-based quasi-2D perovskites are promising materials for nonlinear optics and integrated photonics.}
\end{figure}

\end{document}


\maketitle


\section{Nonlinear optical measurements}
\begin{figure} %
	\centering
	\includegraphics[width=1\textwidth]{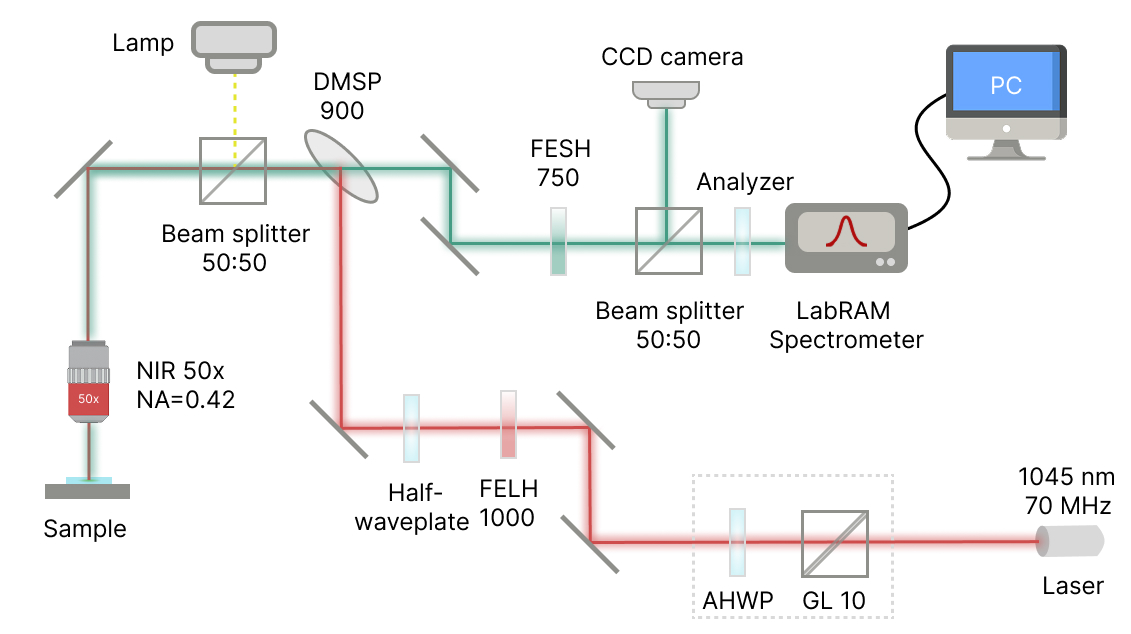}
	\caption{SHG measurements setup.}
	
	\label{fig:setup} 
\end{figure}

All optical experiments were performed on a self-made setup, presented in Figure \ref{fig:setup}. The studied sample was placed on a three-axis stage and rotated at different angles  around its axis. The femtosecond laser emission (1045 nm, 150 fs, 70 MHz) from the Yb-doped laser (TeMa, Avesta Project) was focused onto the sample to a spot size of 3 $\mu$m using 50$\times$ NIR objective (Mitutoyo Plan APO, NA=0.42). The generated second harmonic signal was collected through the same objective and sent to the camera and Horiba LabRam HR spectrometer coupled with a detector (CCD, Andor DU 420A-OE 325). The incident beam and emitted light were separated with a dichroic mirror (DMSP 900, Thorlabs). Fundamental frequency was additionally blocked on the way to the spectrometer by a shortpass filter (FESH 750, Thorlabs). The excitation beam power was adjusted via attenuator (Glan-Taylor prism and half-wavelength plate) and measured with a power meter (Ophir photonics StarBright). Polarization of the incident light was tuned through the rotation of another half-wave plate. The polarization of the SHG signal was selected by an analyzer.
The pump wavelength-dependent measurements were conducted under the femtosecond laser excitation (810-1410 nm, 200 fs, 80 MHz) provided by the optical parametric amplifier (TOPOL-C, Avesta) at a fixed pump power of 10 mW. The laser beam was passing through the dichroic mirror (DMLP 800, Thorlabs) and focused on the sample using the 50$\times$ NIR objective. The generated signal was transferred to the spectrometer (Andor Shamrock 500i).

SHG measurements at different relative humidity (RH) levels were performed using a custom-built chamber equipped with a humidity sensor. The chamber featured a threaded opening for sample placement and a transparent glass window integrated into its top surface. The desired humidity level was established by placing the chamber in an environment with the target RH until the internal humidity stabilized. The chamber was then sealed with Parafilm to preserve these conditions during SHG measurements. The excitation beam was directed onto the sample through the glass window, and the generated SHG signal was collected in reflection geometry. The RH inside the chamber was monitored using the humidity sensor (RH measurement error 0.1 \%).

\section{Nonlinear optical characterization}

\begin{figure} 
	\centering
	\includegraphics[width=0.6\textwidth]{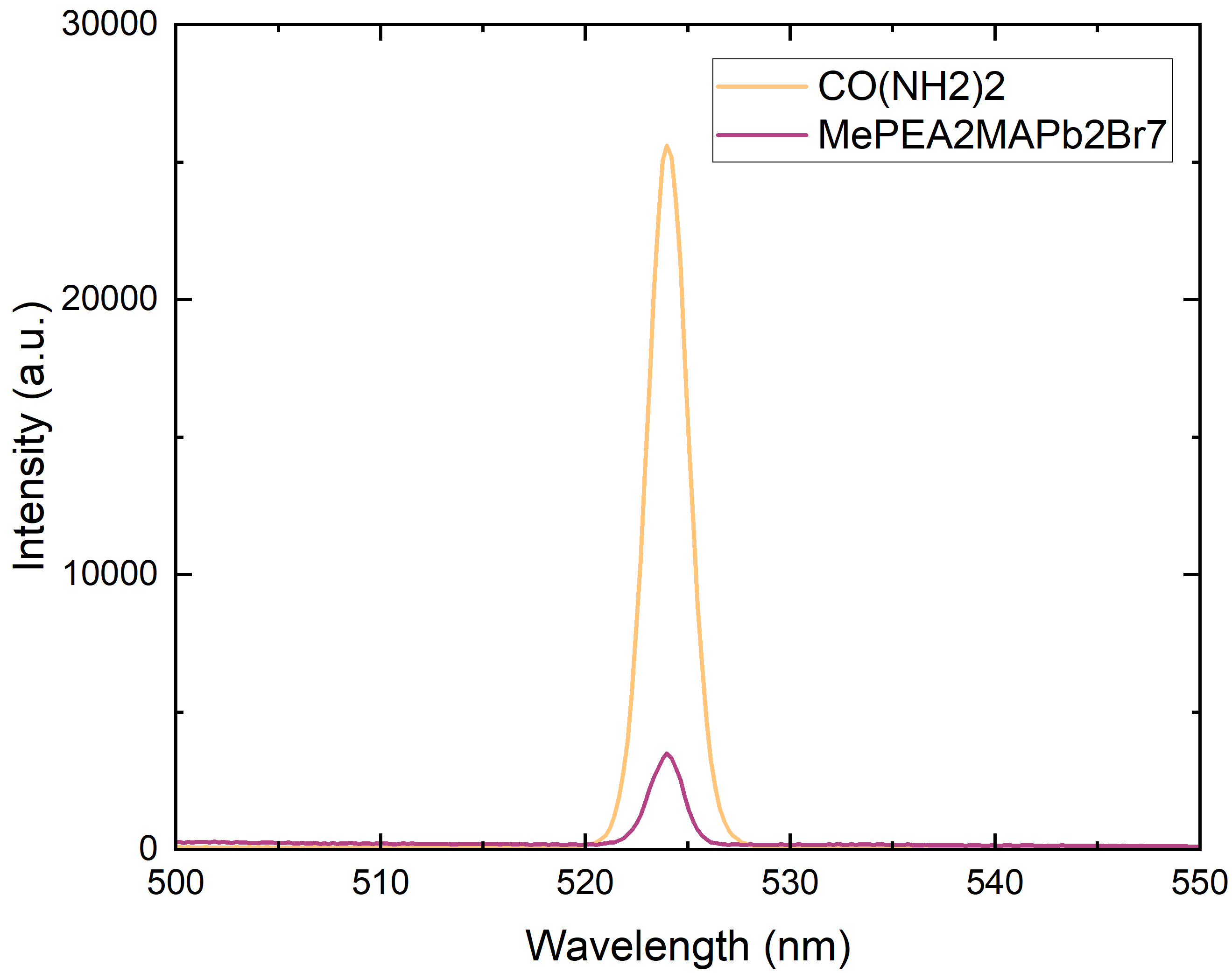}
	 
	\caption{ SHG intensity of MePEA$_2$MAPb$_2$Br$_7$ crystal and crystalline urea at 2.4 mJ/cm$^2$ fluence.}
	\label{fig:eff} 
\end{figure}

Additionally, in this study, we investigated the relative SHG efficiency of MePEA$_2$MAPb$_2$Br$_7$ crystals. Crystalline urea was employed as a standard for SHG \cite{kurtz1968powder}. The SHG efficiency of the MePEA$_2$MAPb$_2$Br$_7$ crystals is found to be 0.15 $\times$ crystalline urea, based on statistics, which were collected for approximately 20 perovskite and urea crystals at varying pumping intensities. The SHG signal from perovskite and urea crystals was measured in the reflection configuration under the same experimental conditions at $\lambda_{\mathrm{exc}} = 1045$ nm (Figure \ref{fig:eff}). The size and shape of the perovskite and urea crystals were selected to be comparable. Based on these measurements, we estimated the second-order susceptibility $\chi^{(2)}$ of MePEA$_2$MAPb$_2$Br$_7$ from the following expression:
\begin{align}
    \chi_S^{(2)}=\chi_R^{(2)}\left|\frac{I_S(2\omega)}{I_R(2\omega)}\right|^{1/2},
    \label{Eq:chi}
\end{align}

where $\chi_R^{(2)}$ is SHG coefficient of
the urea $\chi_R^{(2)}$ $\sim$2.4 pm/V \cite{lin2003mechanism}, and $I_S(2\omega)$ and $I_R(2\omega)$ are the SHG intensities
of the sample and urea reference, respectively. Thus, the absolute second-order SHG coefficient of MePEA$_2$MAPb$_2$Br$_7$ at 1045 nm can be estimated as $\chi_S^{(2)}$ $\sim$0.9 pm/V. This obtained $\chi^{(2)}$ value exceeds that of $\alpha$-quartz ($\chi_{\alpha\text{-quartz}}^{(2)} = 0.67$ pm/V) and is comparable with KDP ($\chi_{\text{KDP}}^{(2)} = 0.88$ pm/V) \cite{fleming2018handbook}. Moreover, it is almost twice as large as the values reported for BA$_2$MAPb$_2$Br$_7$ ($\chi^{(2)} = 0.56$ pm/V) \cite{li2019two}.

\begin{figure} 
	\centering
	\includegraphics[width=0.6\textwidth]{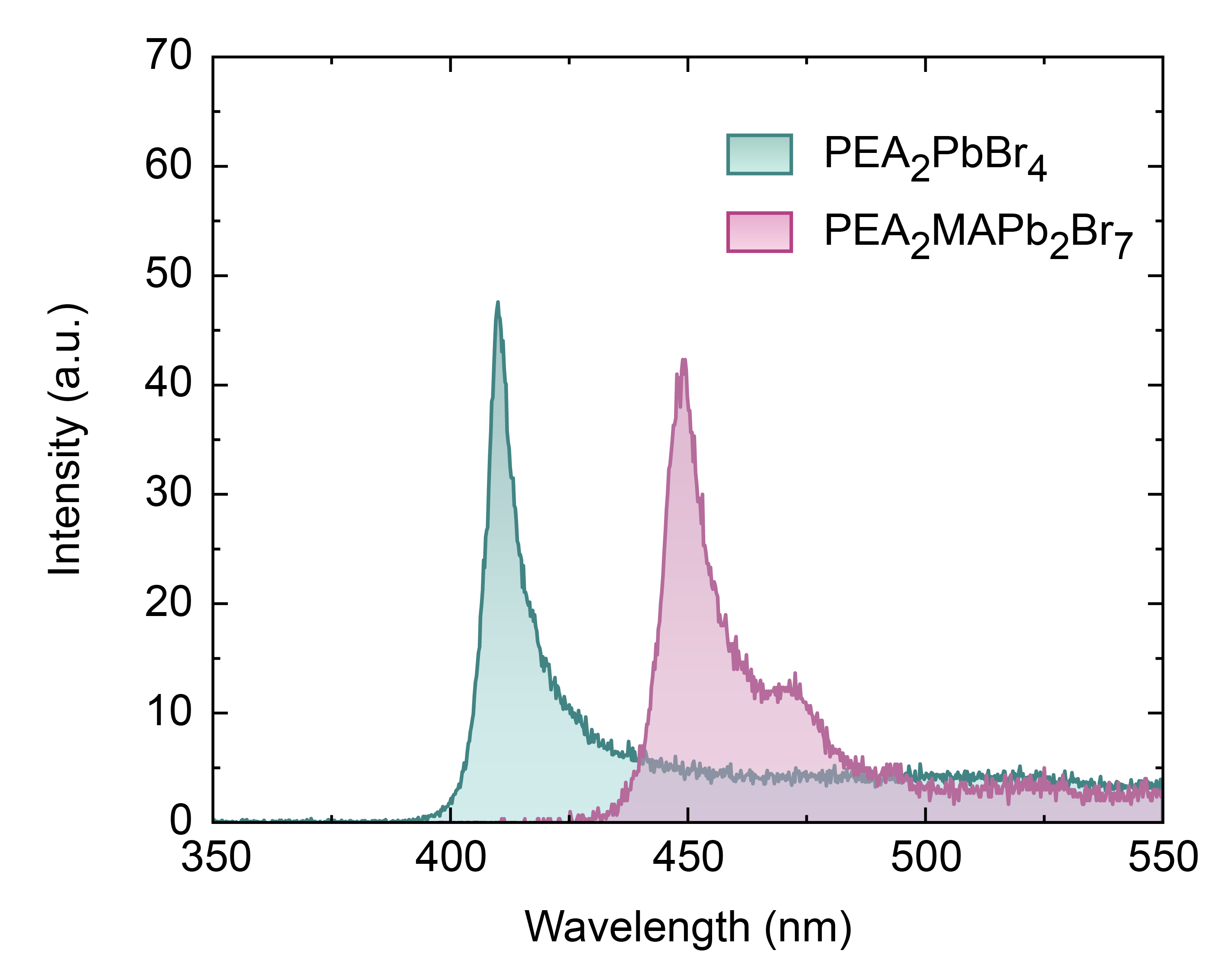}	 
	\caption{ Nonlinear emission spectra of PEA$_2$PbBr$_4$ and PEA$_2$MAPb$_2$Br$_7$ crystals.}
	\label{fig:pea} 
\end{figure}

Figure \ref{fig:pea} shows the mPPL spectra of PEA$_2$PbBr$_4$ and PEA$_2$MAPb$_2$Br$_7$ crystals under 1045-nm excitation at a pump power of approximately 10 mW, with no detectable SHG signal.

\begin{figure} 
	\centering
	\includegraphics[width=1\textwidth]{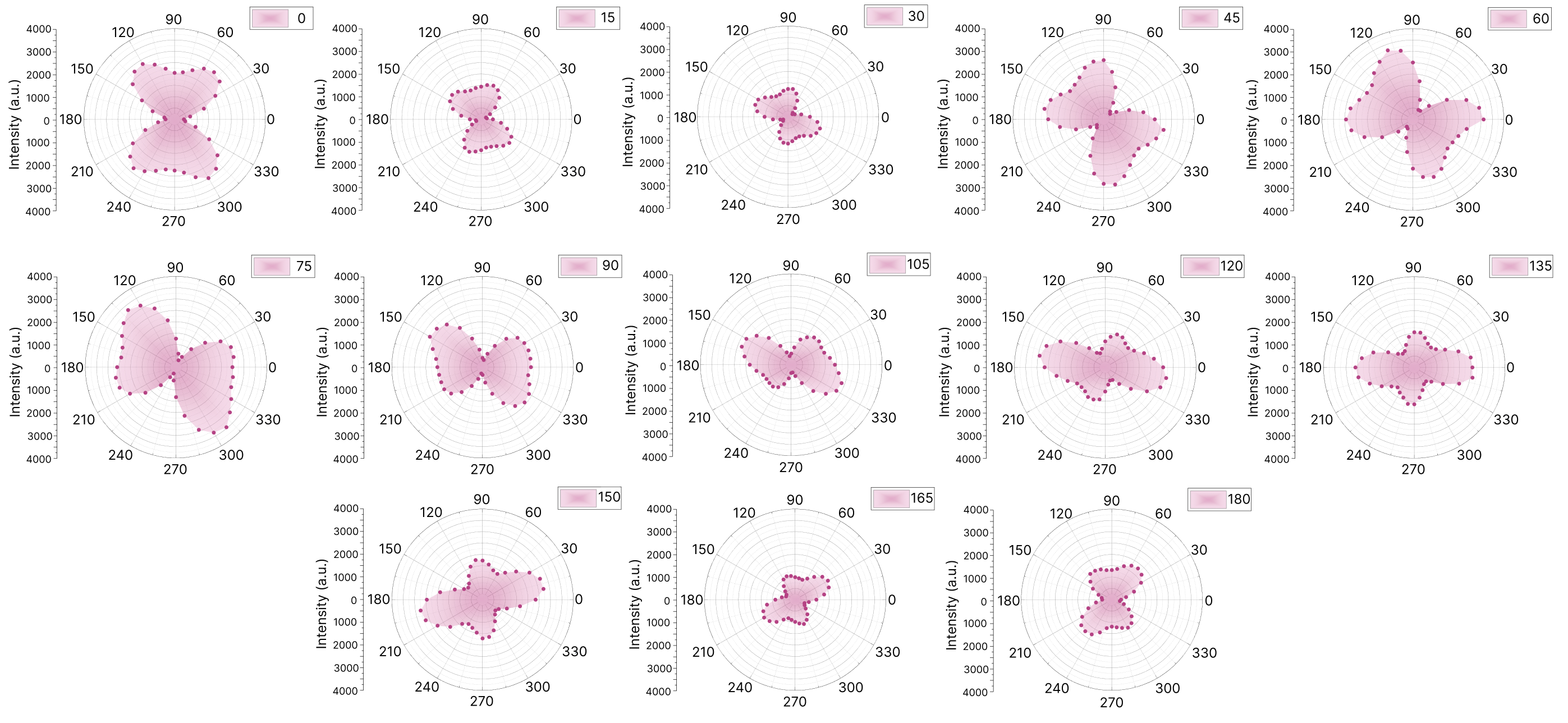}

	\caption{Polar plots of unpolarized SHG intensity as a function of incident polarization angle for different $\theta$ angles, corresponding to substrate rotation in the plane perpendicular to the incident field with a 15$^\circ$ step for MePEA$_2$MAPb$_2$Br$_7$ crystal.}
	
	\label{fig:Polar} 
\end{figure}

\begin{figure} 
	\centering
	\includegraphics[width=1\textwidth]{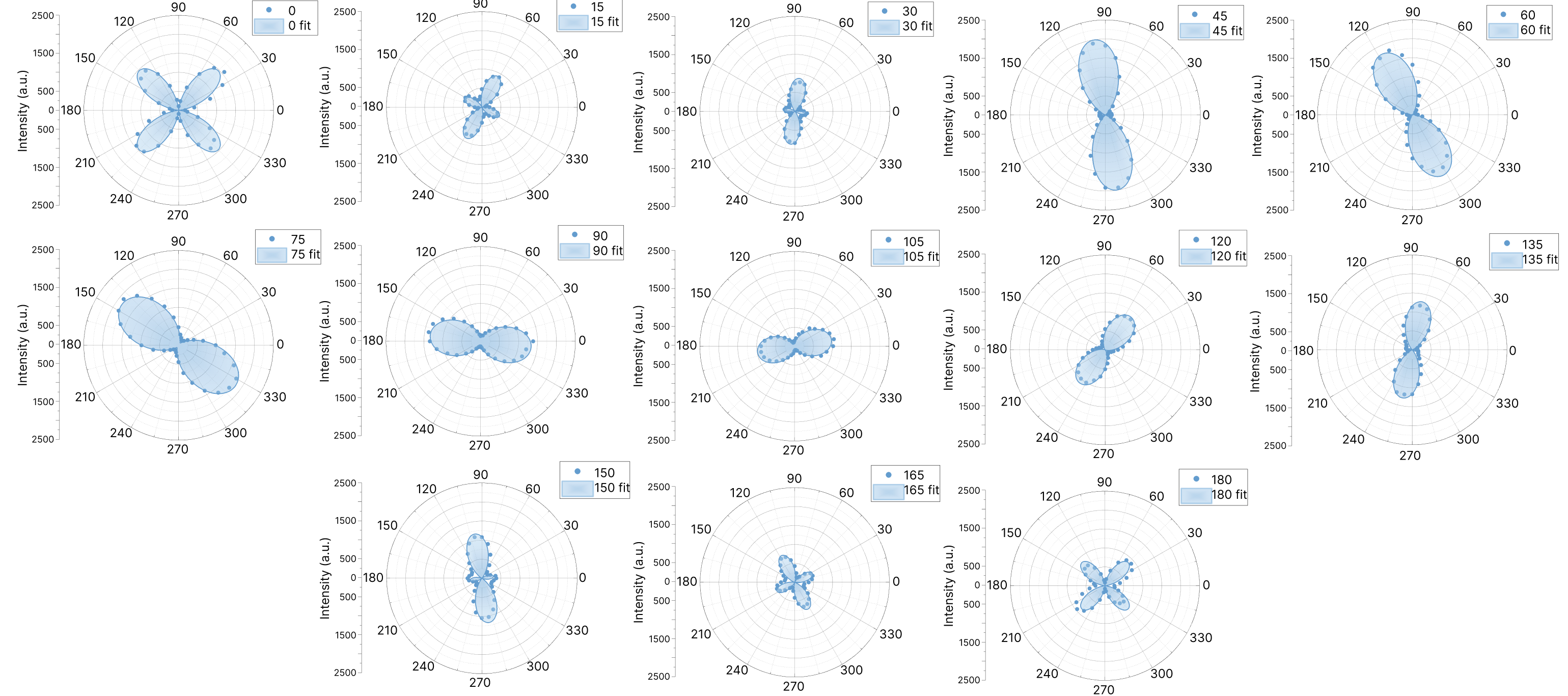}

	\caption{Polar plots of p-polarized SHG intensity as a function of incident polarization angle for different $\theta$ angles, corresponding to substrate rotation in the plane perpendicular to the incident field with a 15$^\circ$ step for MePEA$_2$MAPb$_2$Br$_7$ crystal. Dots represent the experiment data, and solid lines represent fitting.}
	
	\label{fig:PPolar} 
\end{figure}

\begin{figure} 
	\centering
	\includegraphics[width=1\textwidth]{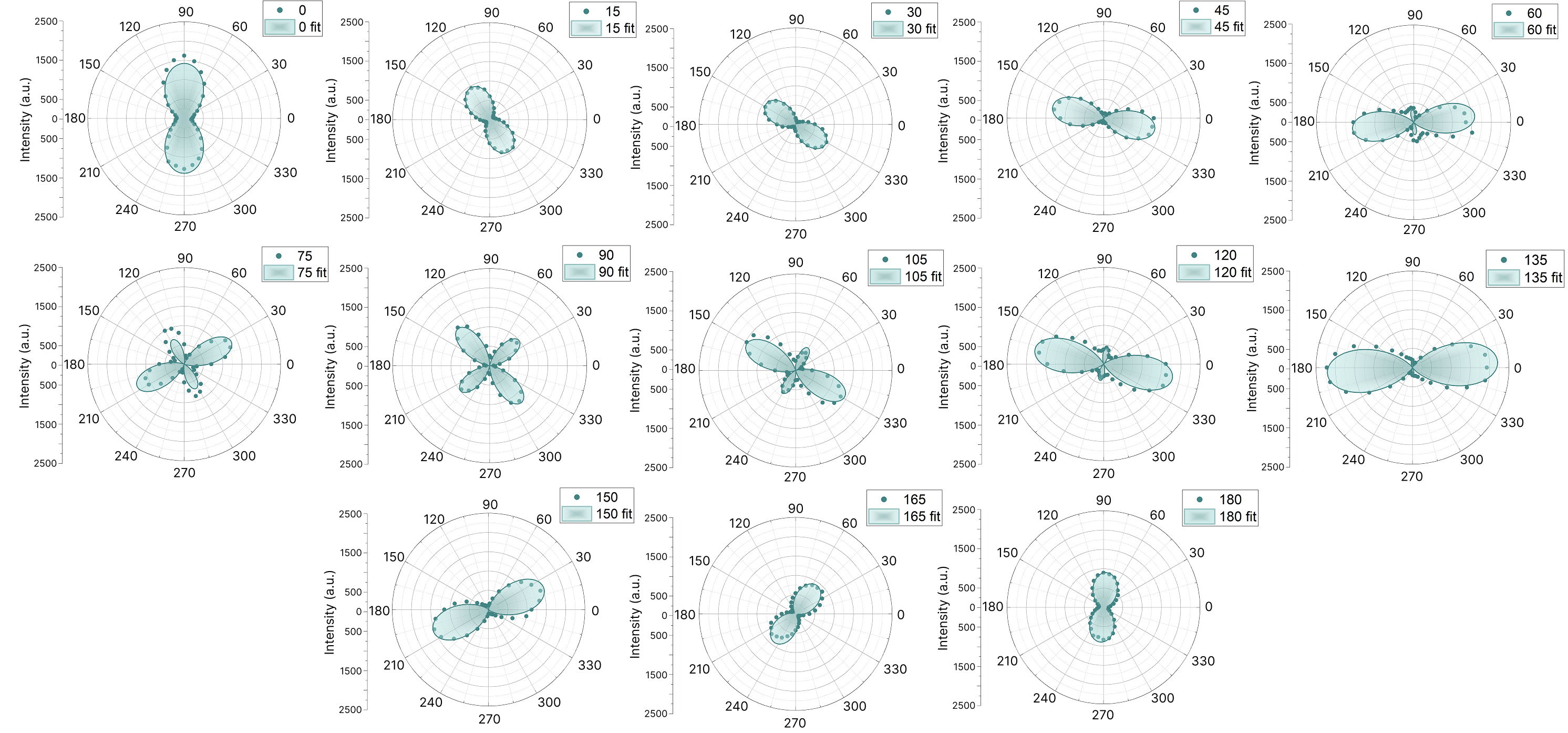}

	\caption{Polar plots of s-polarized SHG intensity as a function of incident polarization angle for different $\theta$ angles, corresponding to substrate rotation in the plane perpendicular to the incident field with a 15$^\circ$ step for MePEA$_2$MAPb$_2$Br$_7$ crystal. Dots represent the experiment data, and solid lines represent fitting.}
	
	\label{fig:sPolar} 
\end{figure}

Figures \ref{fig:Polar}, \ref{fig:PPolar} and \ref{fig:sPolar} present polar plots of the measured SHG response as a function of the sample rotation angle in the plane perpendicular to the incident field, acquired with 15$^\circ$ steps for unpolarized, p-polarized, and s-polarized SHG signals, respectively. The observed evolution of the polar plot shape with sample orientation indicates an anisotropy in the MePEA-based quasi-2D perovskite crystals.

\section{SHG theoretical analysis}

The quasi-2D perovskite crystal MePEA$_2$MAPb$_2$Br$_7$ is characterized by monoclinic symmetry belonging to the $P2$ space group with twofold rotation symmetry about the \textit{c} axis. In the case of the $C_2$ point group, the second-order nonlinear polarization ($P^{2\omega}$) can be written in matrix notation as:

\begin{align}
  \begin{pmatrix}
    P_x^{2\omega}\\P_y^{2\omega}\\P_z^{2\omega}
\end{pmatrix}=
2\varepsilon_0\begin{pmatrix}
0&0&0&d_{14}&0&d_{16}\\d_{21}&d_{22}&d_{23}&0&d_{25}&0\\0&0&0&d_{34}&0&d_{36}  
\end{pmatrix}\begin{pmatrix}
    E_x^\omega E_x^\omega\\E_y^\omega E_y^\omega\\E_z^\omega E_z^\omega\\2E_y^\omega E_z^\omega\\2E_x^\omega E_z^\omega\\2E_x^\omega E_y^\omega
    \end{pmatrix}, 
\label{Eq:1}
\end{align}
where $\varepsilon_0$ is the permittivity of free space and $x$, $y$ and $z$ are the crystal coordinates. Here we have used a convenient contracted notation $d_{im}$ of the second-order nonlinear susceptibility tensor $\chi_{ijk}^{(2)}$, where the first subscript refers to 1: $x$, 2: $y$, 3: $z$ and the second to 1: $xx$, 2: $yy$, 3: $zz$, 4: $yz$, $zy$, 5: $xz$, $zx$, 6: $xy$, $yx$.

\begin{figure} 
	\centering
	\includegraphics[width=0.5\textwidth]{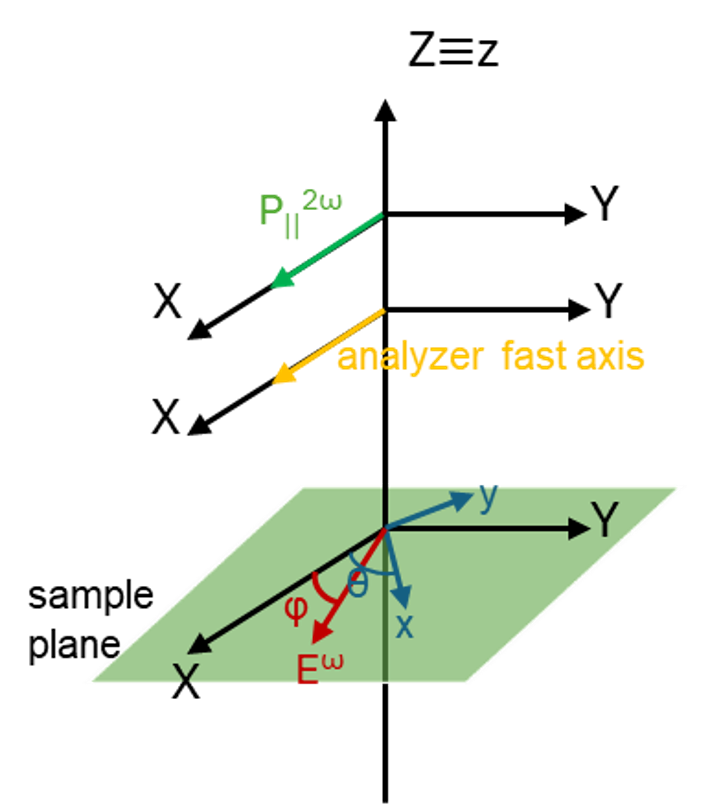}

	\caption{ Coordinate system.}
	
	\label{fig:X} 
\end{figure}

In the experiment, the laser is normally incident on the crystal ($E_z=0$) and linearly polarized in the sample plane, at an angle $\varphi$ with respect to the X laboratory axis (Figure \ref{fig:X}). The orientation of the excitation linear polarization is varied by rotating the half-waveplate. Thus, the incident field in laboratory coordinates can be expressed as $
\begin{pmatrix}
    E_X^{\omega}\\E_Y^{\omega}
\end{pmatrix}
=
\begin{pmatrix}
    E_0\cos{\varphi}\\E_0\sin{\varphi}
\end{pmatrix}
$,
where $E_0$ is the amplitude of the electric field. We then express the excitation field in crystal coordinates by multiplying with the rotation matrix:
\begin{align}
    \begin{pmatrix}
    E_x^{\omega}\\E_y^{\omega}
\end{pmatrix}
= \begin{pmatrix}
    \cos{\left(\theta\right)}&\sin{\left(\theta\right)}\\-\sin{\left(\theta\right)}&\cos{\left(\theta\right)}
\end{pmatrix}\begin{pmatrix}
    E_X^{\omega}\\E_Y^{\omega}
\end{pmatrix}=\begin{pmatrix}
    E_0 \cos{\left(\varphi-\theta\right)}\\E_0\sin{\left(\varphi-\theta\right)}
\end{pmatrix},
\label{Eq:3}
\end{align}
where the crystal \textit{x} axis is taken at an angle $\theta$ from the laboratory \textit{X} axis (Figure \ref{fig:X}).

Considering the propagation of the fundamental and SHG signals along the $Z$-axis, Equation \ref{Eq:1} reduces to:
\begin{align}
    \begin{pmatrix}
    P_x^{2\omega}\\P_y^{2\omega}
\end{pmatrix}=2\varepsilon_0E_0^2
\begin{pmatrix}
    d_{16}\sin{\left(2\left(\varphi-\theta\right)\right)}\\d_{21}\cos^2{\left(\varphi-\theta\right)}+d_{22}\sin^2{\left(\varphi-\theta\right)}
\end{pmatrix}
\label{Eq:4}
\end{align}

We then transform Equation~\ref{Eq:4} back to laboratory coordinates by multiplying with the rotation matrix $\begin{pmatrix}
    \cos{\left(\theta\right)}&-\sin{\left(\theta\right)}\\\sin{\left(\theta\right)}&\cos{\left(\theta\right)}
\end{pmatrix}$, yielding the following expressions for the intensity of the SHG field.

When the analyzer is oriented parallel to the X-axis, the SHG intensity is given by:
\begin{align}
    I_\parallel^{2\omega} 
    &\sim |P_X|^2 \sim \frac{1}{16}( -2\left(d_{21}+d_{22}\right)\sin{\left(\theta\right)}+\left(-2d_{16}+d_{21}- d_{22}\right)\sin{\left(\theta-2\varphi\right)} \notag \\
    &+ \left(-2d_{16}-d_{21}+d_{22}\right)\sin{\left(3\theta-2\varphi \right)})^2    
    \label{Eq:6}
\end{align}

For the case of the analyzer perpendicular to the X-axis:
\begin{align}
    I_\perp^{2\omega} 
    &\sim |P_Y|^2 \sim \frac{1}{16}(2\left(d_{21}+d_{22}\right)\cos{\left(\theta\right)}+\left(-2d_{16}+d_{21}- d_{22}\right)\cos{\left(\theta-2\varphi\right)} \notag \\
    &+ \left(2d_{16}+d_{21}-d_{22}\right)\cos{\left(3\theta-2\varphi \right)})^2
    \label{Eq:7}
\end{align}

The SH intensities $I_\parallel^{2\omega}$ and $I_\perp^{2\omega}$ (or $I_p$ and $I_s$) can also be expressed in terms of dimensionless ratios of the $\chi^{(2)}$ tensor elements in order to fit the experimental data:
\begin{align}
    I_\parallel^{2\omega} 
    &= A( -2\left(B+1\right)\sin{\left(\theta\right)}+\left(-2C+B- 1\right)\sin{\left(\theta-2\varphi\right)} + \left(-2C-B+1\right)\sin{\left(3\theta-2\varphi \right)})^2 \notag\\
    I_\perp^{2\omega}     &=A(2\left(B+1\right)\cos{\left(\theta\right)}+\left(-2C+B-1\right)\cos{\left(\theta-2\varphi\right)} + \left(2C+B-1\right)\cos{\left(3\theta-2\varphi \right)})^2,
    \label{Eq:8}
\end{align}
where A is a multiplication factor and parameters B and C depend on ratios of the second-order susceptibility tensor.

We fit the experimental $I_\parallel^{2\omega}$
and $I_\perp^{2\omega}$ using Equation~\ref{Eq:8} over a wide range of $\theta$ and obtained perfect agreement between the theoretical and measured dependencies (Figures \ref{fig:PPolar} and \ref{fig:sPolar}).

For the case of $\theta = 0$, Equation~\ref{Eq:8} can be further simplified. The total SHG intensity (without an analyzer) then reads $I^{2\omega}= I_\parallel^{2\omega}+I_\perp^{2\omega}\propto \left(2C\sin{\left(2\varphi\right)}\right)^2+(B+1+(B-1)\cos{\left(2\varphi\right)})^2$.The calculated polar dependencies of $I^{2\omega}$, $I_\parallel^{2\omega}$, and $I_\perp^{2\omega}$
for $\theta = 0$ are shown in Figure 3e and 3f of the main text. The extracted relative ratios of the tensor components are estimated as $|d_{22}/d_{21}|$$\sim$2.3 and $|d_{22}/d_{16}|$$\sim$1.16.

\section{SHG after long-term ambient storage}

\begin{figure} 
	\centering
	\includegraphics[width=1.0\textwidth]{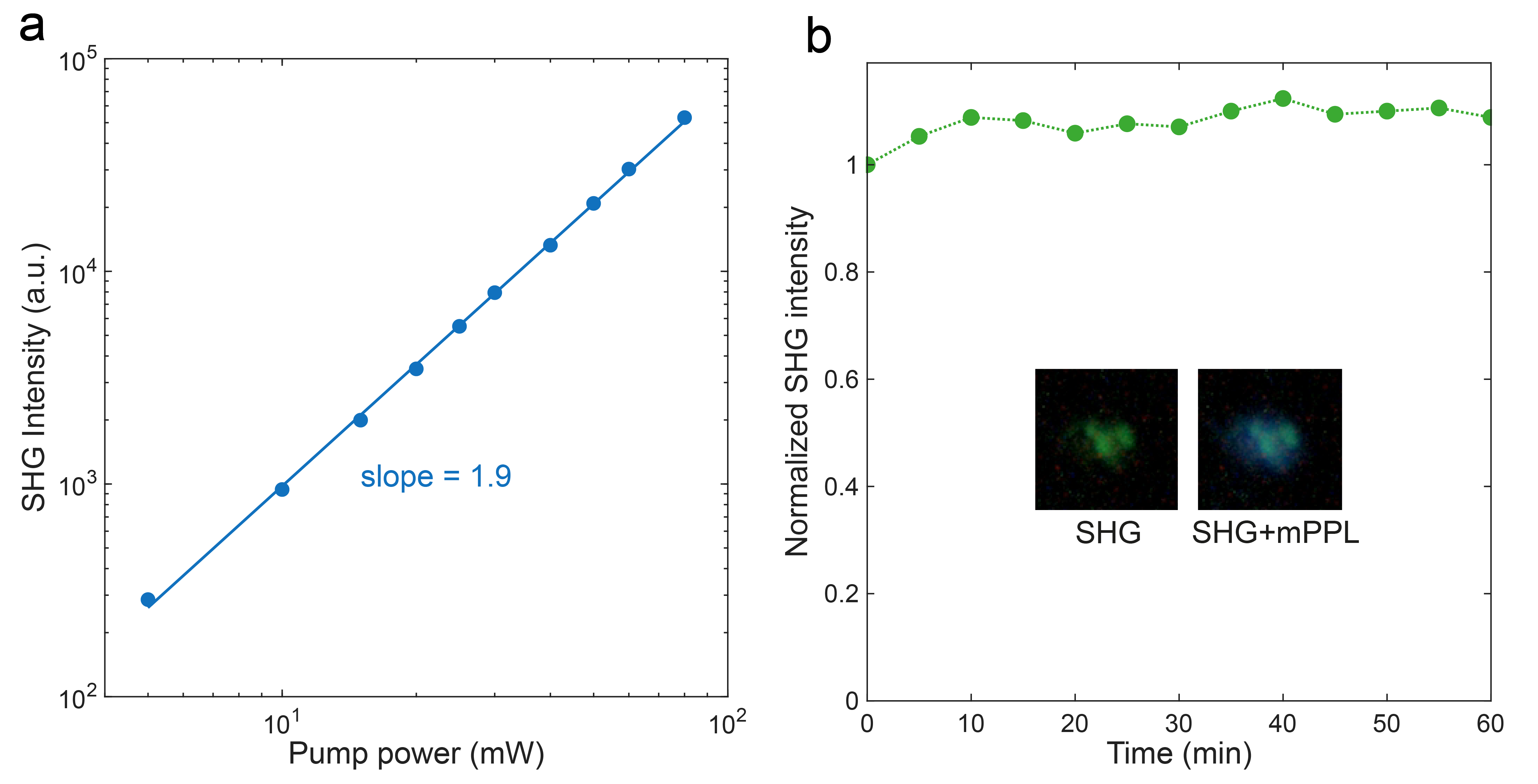}
	
	\caption{(\textbf{a}) Pump-power dependence of the SHG intensity of MePEA$_2$MAPb$_2$Br$_7$ after storage under ambient conditions for 12 months. (\textbf{b}) Temporal evolution of the SHG intensity under continuous excitation at $\lambda_{\mathrm{exc}} = 1045$~nm and a pump intensity of 9.5~GW/cm$^2$, measured at a relative humidity of 50\%. Insets: Optical images of the SHG and combined SHG+mPPL signals.}
	
	\label{fig:Xold} 
\end{figure}

To evaluate the SHG response after long-term storage, a MePEA$_2$MAPb$_2$Br$_7$ crystal stored under ambient conditions for 12 months was investigated. The crystal exhibited a pronounced SHG signal with a nearly quadratic pump-power dependence with fitted slope of 1.9, as shown in Figure \ref{fig:Xold}a. Furthermore, the SHG intensity remained stable during 60~min of continuous femtosecond excitation at 1045~nm with a pump intensity of 9.5~GW/cm$^2$ (Figure~\ref{fig:Xold}b).

\FloatBarrier

\bibliography{3.references}